\documentclass[
    ,final            % use final for the camera ready runs
  ]
  {aipproc}

\layoutstyle{6x9}
\def\psim{\lower.5ex\hbox{$\; \buildrel \propto \over\sim \;$}}
\def\gtrsim{\lower.5ex\hbox{$\; \buildrel > \over\sim \;$}}
\def\lesssim{\lower.5ex\hbox{$\; \buildrel < \over\sim \;$}}
\def\e{{\epsilon}}

\def\apj{{\it Astrophys.\ J.}}
\def\apjl{{\it Astrophys.\ J.\ Lett.}}
\def\nat{{\it Nature}}
\def\mnras{{\it Monthly Not.\ Roy.\ Astr.\ Soc.}}
\def\e{{\epsilon}}

\begin{document}

\title{First Light on GRBs with {\it Fermi}}


\author{Charles D. Dermer\\
{\small on behalf of the {\it Fermi} Collaboration}}{
  address={Code 7653, Naval Research Laboratory, 4555 Overlook Ave.\ SW, Washington, DC 20375-5352 USA}
}

%\author{et al. }{
%  address={address}
%}

\begin{abstract}
{\it Fermi} LAT (Large Area Telescope) and GBM (Gamma ray Burst Monitor) observations of GRBs 
are briefly reviewed, keeping in mind EGRET 
expectations. Using $\gamma\gamma$ constraints on outflow Lorentz factors,
leptonic models are pitted against hadronic models, and found to be energetically favored.
Interpretation of the {\it Fermi} data on GRBs 
helps establish whether 
GRBs accelerate cosmic rays, including those reaching $\approx 10^{20}$ eV.
\end{abstract}

\maketitle

\section{Introduction}

The effective lifetime for GRB studies 
using EGRET's spark chamber on the {\it Compton Gamma-ray Observatory}
ended $\approx 4.5$ yrs into 
mission, after 1996  \cite{kur97}.\footnote{Very bright GRBs like  
GRB 990123\cite{bri99} could still be detected far off the  
COMPTEL and OSSE axes while making a signal in the Energetic Gamma Ray
Experiment Telescope's Total Absorption Shower Counter.
EGRET TASC and BATSE data  were used to made the discovery of the additional hard component in 
GRB 941017  \cite{gon03}.}
The depletion of spark-chamber gas was mitigated through 
the introduction of a narrow-field
mode suitable for pointed observations. This made 
the chance of catching a GRB, 
proportional to EGRET's field-of-view (FoV), too improbable
without rapdi, automated slewing, which was not possible for CGRO. Consequently
EGRET only detected a total of five spark-chamber GRBs, all 
early in the mission  \cite{din95}. These are GRB 910503, GRB 910601, 
the superbowl GRB 930131, the famous long-lived GRB 940217 \cite{hur94}, and GRB 940301. 
In the wide-field mode, EGRET was sensitive to 
$\approx 1/25^{th}$ of the full sky, 
which is $\approx 1/5^{th}$ as large as the FoV of {\it Fermi} \cite{atw09}.

Since the {\it Fermi Gamma-ray Space Telescope} science operations began, from early August 2008 through calendar year 2009, 
13 GRBs were reported as significantly detected in the LAT by the 
{\it Fermi} Collaboration.\footnote{See {\it Fermi}.gsfc.nasa.gov/ssc/observations/types/grbs } All LAT GRBs are also 
GBM GRBs and comprise the brighter GBM GRBs, as already expected
from a comparison between EGRET and BATSE GRBs in terms of fluence \cite{led09}.
The 13 {\it Fermi} LAT GRBs include 11 long GRBs
and 2 short bursts, namely
GRB 081024B and GRB 090510 ($z = 0.903$). 
The most studied---because they are brightest---GRBs are GRB 090902B ($z = 1.822$) \cite{abd09_grb090902b}, 
which provides the 
first strong evidence for a hard spectral component in long GRBs;
GRB 080916C ($z = 4.35$) 
\cite{abd09_grb080916c}, the first bright long GRB; and GRB 090926A ($z = 2.106$), a burst with a narrow spike
from the lowest to highest energies in an SED that requires both 
a Band function and a hard power-law component to fit. 
As discussed at this conference by M.\ Ohno and T.\ Uehara,
%  \cite{abd09_grb090926a}, 
GRB 090926A also reveals an extraordinary spectral softening at 
$\gtrsim 1$ GeV in its time-integrated spectrum when the hard LAT spectral 
component is bright. All these GRBs were
more fluent than the fiducial fluence 
$\Phi_{fid} = 10^{-4}$ erg cm$^{-2}$ in the  20 -- 2000 keV range 
that GBM measures
(Fig.\ 1). 
The bright, short GRB 090510, with $\Phi \equiv \Phi(20$ -- 2000 keV) $\cong 10^{-5}$ erg cm$^{-2}$, 
also shows (like GRB 090902B and GRB 090926A)
 a distinct high-energy power-law spectral component in addition to a Band component  
\cite{ack10_grb090510}. 
Its short duration, large distance, and the detection of a 
31 GeV photon permit strict tests on quantum gravity theories predicting a 
dependence of the speed of light in vacuo that is linear with energy
 \cite{abd09_grb090510}.

\begin{figure}[t]
 \hskip0.8in \includegraphics[height=.36\textheight]{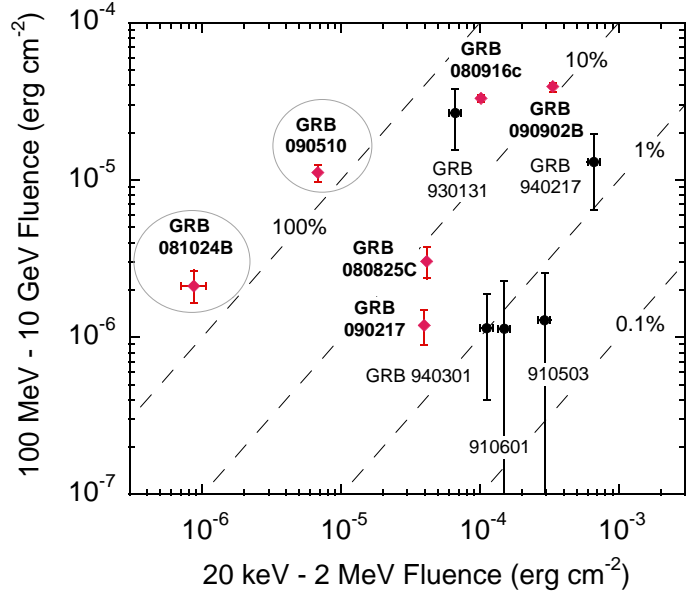}
  \caption{Fluence-fluence diagram showing 6 {\it Fermi} [14] GRBs (red data points) 
and 5 EGRET spark-chamber [7] GRBs (black data points). The EGRET fluence is measured from 100 MeV to 5 GeV, whereas the {\it Fermi} LAT fluence
is measured from 100 MeV to 10 GeV. Short hard GRBs are circled.}
\end{figure}

Besides these notable GRBs are the less well-known and also less fluent long duration
GRBs 090323 ($z = 3.57$), 090626 and 090328 ($z = 0.736$), with $\Phi \approx 10^{-4}$ erg cm$^{-2}$,
the widely off-LAT-axis GRB 081215A, the first LAT GRB 080825C \cite{abd09_grb080825c}, the 
unusual GRB 090217 \cite{ack10_grb090217a} showing none of the typical properties of LAT GRBs,
and the undistinguished LAT GRBs 091003A and GRB 091031.  
The weakest fluence GRB of the sample is the first short GRB detected at LAT energies, GRB 081024B \cite{abd10_grb081024b},
with $\Phi\approx 4\times 10^{-7}$ erg cm$^{-2}$. The weakness of this 
GRB could be related to the high $E_{pk}\approx 2$ -- 3 MeV of its Band-function component, 
but the time-averaged $E_{pk}\cong 4$ MeV for GRB 090510 between 0.5 and 1  s after trigger is
even higher \cite{ack10_grb090510}.

 For those GBM GRBs occurring within the LAT FoV,
detection of GRBs with the LAT is almost guaranteed when $\Phi\gtrsim 10^{-4}$ erg cm$^{-2}$. 
The detection rate slips to less than 50\% when $\Phi\approx 3\times 10^{-5}$ erg cm$^{-2}$, 
and becomes highly improbable for $\Phi\lesssim 10^{-5}$ erg cm$^{-2}$.
This behavior undoubtedly reflects a distribution in the ratios of $\gtrsim 100$ MeV LAT to 
GBM  energy fluence \cite{gpw10}. 

Some implications from first results on GRBs from the {\it Fermi} Gamma ray Space Telescope 
are considered in this paper.
Besides presenting brief additional description about LAT and GBM observations of the 13 {\it Fermi} LAT GRBs, 
we consider the question of leptonic vs.\ hadronic origin of the 
high-energy $\gamma$ rays from GRBs, and whether hadronic signatures are detected
in the high-energy spectra of GRBs.

\section{LAT and GBM Observations of GRBs}

The Gamma-ray Large Area Space Telescope was launched 11 June 2008, and went into science operations
two months later, in early August 2008, shortly before being renamed after Enrico {\it Fermi}.
In these 16 months, $\lesssim 1$ GRB per month was detected with the {\it Fermi} LAT,
 or $\approx 9$ GRB/year, with LAT detecting short GRBs 
at $\approx 10$ -- 20\% of the rate of long GRBs. GRBs are detected with the GBM at a rate of 
250 GRB/yr, or $\approx 500$ GRB/yr (full-sky). 
When corrected for FoV, EGRET detected $\approx 25$ GRB/year (full sky), while the {\it Fermi} LAT detects 
$\approx 50$ GRB/yr (full sky). Given the much larger effective area 
of {\it Fermi} than EGRET, by a factor $\approx 6 ~[\approx (8000$ -- $9000 {\rm ~cm}^{2})/(1200$ -- $1500 $ cm$^{2}$)], 
this small rate increase is something of a surprise, compounded by the ongoing sparse period of
{\it Fermi} LAT detections of GRBs in the first half of 2010. Part of this difference
is the stronger detection criteria of {\it Fermi} LAT than EGRET. 
But an improvement in flux sensitivity by a factor $\approx 6$, with an accompanying rate increase by 
only a factor $\approx 2$ -- 3 suggests that LAT GRBs are being 
sampled in a portion of their $\log N-\log \Phi$ distibution that is flattened by 
cosmological effects. This is consistent with the known redshifts of LAT GRBs, which 
range from $\approx 0.7$ to $z = 4.35$, with a very rough average redshift of $\langle z \rangle=2$ for long
GRBs and $\langle z \rangle \approx 1$ for short GRBs (based only on GRB 090510). If typical,
both classes of GeV-emitting GRBs 
would be subject to strong cosmological effects on the fluence and flux distributions. 

Pre-{\it Fermi} estimates of the rate of LAT detections are given in Refs.\ \cite{led09,ban09}.

\subsection{Fluence-fluence diagram}

Fig.\ 1 shows the fluence-fluence diagram for the 5 EGRET spark-chamber \cite{led09} 
and for 6 {\it Fermi} \cite{abd10_grb081024b} GRBs
(values for the other 7 {\it Fermi} LAT GRBs, which tend to be 
the dimmer LAT GRBs, await final {\it Fermi} analysis). Most GBM GRBs
have $\Phi \lesssim 10^{-5}$ erg cm$^{-2}$, 
and are only rarely detected with the LAT. 
Because of the small number of LAT GRBs, it is not yet 
clear whether there is a 
systematic difference between fluence ratios of EGRET and 
{\it Fermi} LAT-detected GRBs. The weakest {\it Fermi} LAT
GRBs in terms of GBM fluence are both short duration GRBs. 
This could indicate a preference for short GRBs to have a larger ratio 
of LAT to GBM fluences than long GRBs, depending on 
possible triggering biases, e.g., increased LAT 
background for long GRBs. 

The apparent isotropic energies ${\cal E}_{iso}$ of GBM and LAT emission for LAT GRBs
with known redshifts are in several cases $\gtrsim 10^{54}$
erg. For GRB 080916C, ${\cal E}_{iso} \approx 10^{55}$ erg. The LAT GRBs tend to 
have the largest energies of all measured GRBs, and as a result are good for radio studies
\cite{cen10}.

\subsection{{\it Fermi} LAT GRB Phenomenology}

From the first 13 GRBs that have been detected with the LAT, some distinct and 
unexpected behaviors have been identified. In rough order of decreasing significance,
they are:
 
\begin{itemize}

\item Extended (long-lived) LAT (100 MeV -- GeV) emission compared to the GBM 
(20 keV -- 2 MeV) emission, known already from EGRET observations, especially GRB 940217 \cite{hur94}. 

\item Delayed onset of the LAT emission compared to the GBM emission in both long
and short classes of GRBs.

\item Power-law temporal decay profiles of the LAT extended emission, decaying typically $\propto t^{-1.5}$ \cite{ghi10}. 

\item Appearance of separate power-law spectral components with photon number index harder than $-2$.

\item Delayed onset of the lowest energy GBM emission at $\approx 10$ keV, seen for example 
in GRB 090902B and GRB 090926A.

\item Quasi-thermal Band function components with steep Band $\beta$ 
found, e.g., in GRB 090902B at $E\gtrsim 1$ MeV \cite{abd09_grb090902b}.  

\end{itemize}

The onsets of the $> 100$ MeV emission appear to be delayed by $\sim 0.1 t_{90}$
compared to the 100 keV -- MeV emission (with $t_{90}$ measured, e.g., in the 50 -- 300 keV GBM/BATSE range). 
This is 
one of the key and unanticipated results on GRBs from {\it Fermi},
and it appears to operate equally for both the long- and  
short-duration LAT GRBs. There have as yet been no LAT detections of members of 
the low-luminosity/sub-energetic class of GRBs that includes GRB 980425
and GRB 030329, nor have any X-ray flashes or X-ray luminous GRBs been detected
with the LAT. Because GBM's primary triggering modes are similar to BATSE, 
high $E_{pk}$, relatively low-$z$ GRBs (compared to {\it Swift}) are more likely to 
be detected. The differences and similarities between the redshift distributions of LAT,
BATSE/Beppo-SAX, and {\it Swift} GRBs deserve a separate study.

\section{Theoretical Implications}

\subsection{Minimum Bulk Lorentz Factor and Magnetic Field}

Emission from bulk magnetized plasma in relativistic motion
provides the best explanation for the large apparent isotropic 
luminosities   $L_{iso}$ and energy releases ${\cal E}_{iso}$,
from GRBs at cosmological distances. 
Two crucial quantities for modeling the GRB SEDs are
 the bulk Lorentz factor $\Gamma$ of the outflow, and the magnetic
field $B^\prime$, which depends on the comoving 
emission-region size scale $R^\prime$ through the observed variability timescale 
$t_{\rm var}\gtrsim (1+z) R^\prime/\Gamma c$.

Combined {\it Fermi} LAT and GBM observations give the most reliable 
measurements of the minimum bulk outflow Lorentz factor $\Gamma_{min}$
through $\gamma\gamma$ opacity arguments.
It is simple to derive $\Gamma_{min}$ in a blast-wave formulation,
noting that the internal photon energy density 
\begin{equation}
u^\prime_\gamma \approx {4\pi d_L^2 \Phi\over \Gamma^2 4\pi R^2 c} \approx
{(1+z)^2d_L^2\Phi\over \Gamma^6 c^3 t_{\rm var}^2}\;,
\label{eq1}
\end{equation}
using $R\approx \Gamma^2 ct_{\rm var}/(1+z)$. The optical depth for 
$\gamma\gamma\rightarrow e^+e^-$ processes is $\tau_{\gamma\gamma}(\epsilon_1^\prime)
\cong R^\prime \sigma_{\rm T} (\epsilon_1^\prime/2)u^\prime_\gamma(2/\epsilon_1^\prime)/(m_ec^2)$, where 
$R^\prime = R/\Gamma$ and $\epsilon^\prime = 2/\epsilon_1^\prime$ from the threshold 
condition. The condition $\tau_{\gamma\gamma}(\epsilon_1^\prime)< 1$ 
with the relation $\Gamma\epsilon^\prime_1/(1+z) = \epsilon_1$ 
implies 
\begin{equation}
\Gamma \gtrsim\Gamma_{min} \cong \left[ {\sigma_{\rm T} d_L^2 (1+z)^2 
f_{\hat \e}\e_1\over 4t_{\rm var} m_ec^4}\right]^{1/6}\;,\;
\hat \e = {2\Gamma^2\over (1+z)^2\e_1}\;
\label{eq2}
\end{equation}
\cite{ack10_grb090510}.
Here $f_\e $ is the $\nu F_\nu$ flux at photon energy $m_ec^2
\epsilon$, which is evaluated at $\e = \hat \epsilon$ due to 
the peaking of the $\gamma\gamma$ cross section near
threshold. 

Large values of $\Gamma_{min}\sim 10^3$ are deduced from 
the detection of multi-GeV photons with the LAT through detailed calculations or using 
simple estimates with formulas such as eq.\ (\ref{eq2}). Values of $\Gamma_{min} \approx 1280, 1000,$ 
and 1100 are found for GRB 090510 ($E_{max} \cong$ 31 GeV; $t_{\rm var} \cong 0.01$s \cite{ack10_grb090510}), 
GRB 090902B ($E_{max} \cong$ 11 GeV; $t_{\rm var} \cong 0.05$s \cite{abd09_grb090902b}), 
and GRB 080916C ($E_{max} \cong$ 3 GeV; $t_{\rm var} \cong 0.5$s \cite{abd09_grb080916c}), respectively.

Introducing $\epsilon_B$ and $\epsilon_e$ factors for the fraction of 
internal energy in magnetic field and electrons, respectively,
implies from eq.\ (\ref{eq1}) that for GRB 080916C,
\begin{equation}
B^\prime({\rm kG}) \approx 2{\sqrt{\epsilon_B(\Phi/10^{-5}{\rm~erg~cm}^{-2}{\rm~s}^{-1})/\epsilon_e}
\over \Gamma_3^3 t_{\rm var}({\rm s}) }\lesssim 3 \sqrt{{\epsilon_B\over \epsilon_e}}
\;[E_1(10{\rm~GeV})t_{\rm var}({\rm s})]^{-1/2}\;,
\label{eq3}
\end{equation}
where the final expression follows \cite{rdf10} by substituting eq.\ (\ref{eq2})
into eq.\ (\ref{eq1}), and  $\Gamma = 10^3\Gamma_3$.

This can be compared with an equipartition estimate of the magnetic field assuming
that the MeV -- GeV radiation is nonthermal synchrotron radiation, giving
\begin{equation}
B_{eq}^\prime({\rm kG}) \approx 0.058\; {d_{28}^{4/7} f_{-5}^{2/7} \Lambda_2^{2/7} (1+z)^{5/7}
\over  [t_{\rm var}({\rm s})]^{6/7} \Gamma_3^{13/7} \epsilon^{1/7} } \approx 3\;
{f_{-5}^{2/7} \Lambda_2^{2/7}
\over  [t_{\rm var}({\rm 0.1~s})]^{6/7} \Gamma_3^{13/7} \epsilon^{1/7} },
\label{eq4}
\end{equation}
where $10^{-5}f_{-5}$ erg cm$^{-2}$ s$^{-1}$ is the $\nu F_\nu$
flux, $\Lambda \equiv (1+\zeta_{pe}\ln(\epsilon_2/\epsilon_1 ) = 100\Lambda_2$ is the product of the ratio 
$\zeta_{pe}$ of proton to electron energy and a bolometric factor, 
and the last expression applies to GRB 080916C. Here $\epsilon_1$
and $\epsilon_2$ bracket the $\nu F_\nu \propto \nu^{1/2}$ portion of the
SED from electron synchrotron radiation. The coincidence of these two numbers, 
in spite of depending separately on $\epsilon_1$ and $\zeta_{pe}$, gives some confidence that 
$B^\prime_{eq}\sim 1$ -- 10 kG for GRB 080916C, with an energy and jet-power 
penalty $\propto (B^\prime/B^\prime_{eq})^2$ for larger values of $B^\prime$.

\subsection{Afterglow Synchrotron Models}

The problem is to determine which part of the $\nu F_\nu$ spectrum is made by electron 
synchrotron radiation. A well-studied model \cite{dcm00} for the hard X-ray and $\gamma$ radiation
assumes that the BATSE/GBM 
MeV radiation is  electron synchrotron radiation, 
and the $\gtrsim 100$ MeV - GeV emission is synchrotron self-Compton $\gamma$ rays 
made by the decelerating blast wave in 
the early afterglow phase. This model is not favored by the {\it Fermi} data. 
The early appearance of a hard component, the lack of a transition spectral episode 
between dominant synchrotron and self-Compton components at LAT energies, and the line-of-death 
problem \cite{pre98} cannot be easily explained within the framework of such a model, as 
noted already in the interpretation of GRB 941017 \cite{gon03}.

An interesting alternate approach is to assume that the {\it Fermi} LAT emission 
is electron synchrotron radiation made at an external shock formed by outflowing
plasma. In this picture, the time when the LAT flux is brightest corresponds to 
the deceleration time, and is consistent with $\Gamma >> \Gamma_{min}$.  One approach is to suppose that the blast wave decelerates
adiabatically in a uniform surrounding medium \cite{kb09}, with closure relation
$\nu F_\nu \propto t^{(2-3p)/4} \nu^{(2-p)/2}$, where $p$ is the electron injection 
index. A value of $p\approx 2.5$ gives a reasonable fit to the data.
Another regime to consider is a radiative GRB blast wave \cite{ghi10}, where the 
comparable closure relation is  $\nu F_\nu \propto t^{(2-6p)/7} \nu^{(2-p)/2}$, with
$p\approx 2$ giving a plausible fit to the data. The adiabatic model 
requires unusually low densities and magnetic fields for GRB 080916C, and the 
radiative model supposes pair loading can help achieve strong cooling. 

Alternate leptonic models for {\it Fermi} LAT GRBs 
include photospheric models with the 
photospheric emission passing through shocked
plasma in the colliding shells or external shocks
of the GRB outflow \cite{twm10}. A joint {\it Fermi}-{\it Swift} paper 
examines leptonic afterglow models for GRB 090510 \cite{pas09}. 
Such models are considered 
in more detail by P.\ M\'esz\'aros at this conference.

\subsection{Hadronic Models and GRBs as UHECR Sources}

A proton synchrotron model was proposed to explain the origin 
of the delayed onset of the LAT emission in GRB 080916C \cite{rdf10}  
as a result of the time to energize, accumulate and cool protons 
in the swept-up material at an external shock. The difficulty of 
this model is the required large energy in protons 
and magnetic fields for efficient proton-synchrotron
radiation, which means that the apparent isotropic power exceeds
\begin{equation}
 L_B \cong  R^2 c\Gamma^2 B^{\prime 2}/2 \cong 2\times 10^{58}\Gamma_3^{16/3}t^{2/3}_{syn}({\rm s})
E_{100}^{-2/3}  \;
{\rm erg~s}^{-1},
\label{eq5}
\end{equation}
where $100 E_{100}$ MeV is the proton-synchrotron cooling frequency \cite{wan09,rdf10}.
Only if the opening angle of the jet is very narrow and the $\Gamma$ factor
is lower than estimated through the simple $\gamma\gamma$ arguments, is the proton
synchrotron model energetically reasonable. Applying this expression to GRB 090510 gives
an absolute energy output of ${\cal E}_{abs} \cong 10^{53} \Gamma_3^{16/3} t^{5/3}_{onset}({\rm 0.1 s})
 E_{100}^{-2/3} [\theta_j({\rm deg})]^2$ erg. Lower $\Gamma$ factors arising from
dynamic and particle cooling effects \cite{gra08} could ameliorate the large implied energies.
A hybrid lepton-proton synchrotron model can explain the afterglow 
LAT and {\it Swift} light curves for GRB 090510 with very small values of $\epsilon_e$
and large values of $\epsilon_B$ \cite{raz10}.

\begin{figure}[t]
\hskip0.8in  \includegraphics[height=.36\textheight]{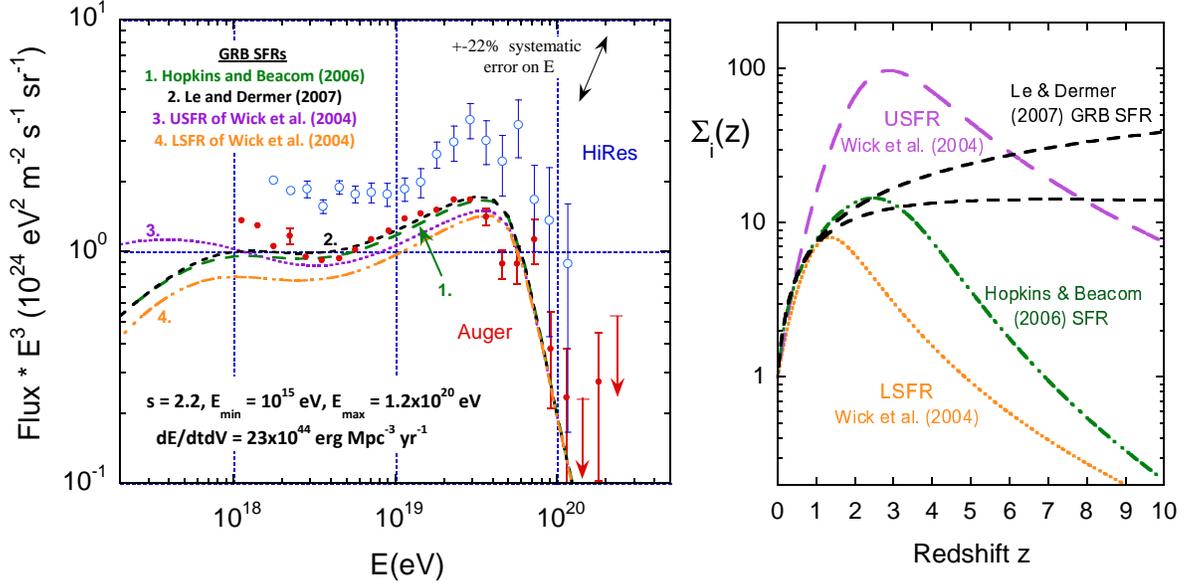}
  \caption{Left panel shows fits of models of UHECRs from GRBs to Auger data. UHECR protons are 
injected between $10^{15}$ eV ($\approx 10^6\Gamma_3^2m_pc^2$) and an exponentially cutoff energy of
$1.2\times 10^{20}$ eV with a $-2.2$ number spectrum for different GRB star formation 
rate functions shown in the right panel. The local UHECR emissivity from GRBs is normalized
to $23\times 10^{44}$ erg Mpc$^{-3}$ yr$^{-1}$.
 }
\end{figure}

The photopion efficiency for protons interacting with photons 
with energies near the peak photon energy $E_{pk} = \epsilon_{pk} m_ec^2$ is given 
by \cite{wb97,mur06}
\begin{equation}
\eta_{p\gamma}(E_p^{pk}) \cong {K_{p\gamma}\sigma_{p\gamma}d_L^2 f_{\e_{pk}}
\over m_ec^4 \Gamma^4 t_{\rm var}\epsilon_{pk}}\cong 0.03\;{f_{-5}\over \Gamma_3^4 (t_{\rm var}/0.01{\rm ~s})\epsilon_{pk}}
\label{eq7}
\end{equation}
where $K_{p\gamma}\sigma_{p\gamma}\cong 70\mu$b and 
the final expression applies to GRB 090510.
The escaping energy of protons
that interact with photons in the 
blast wave at $\e^\prime \approx \e^\prime_{pk}$ is 
$E_p^{pk} \cong 400 m_pc^2 \Gamma^2/(1+z)\epsilon_{pk} 
\cong 2\times 10^{17} \Gamma_3^2/\epsilon_{pk}$ eV.
The efficiency varies $\propto (E_p/E_p^{pk})^{-1-\beta }$ at low
energies (where the protons are interacting with the higher
energy portion of the Band spectrum), and 
$\propto (E_p/E_p^{pk})^{-1-\alpha }$ when $E_p > E_p^{pk}$ in 
terms of of Band $\alpha$ and $\beta$. 
As considered in more detail at this conference by K.\ Asano,
the large $\Gamma$ factors inferred for {\it Fermi} LAT GRBs are 
unfavorable for $\sim$PeV neutrino production or the generation
of strong cascade $\gamma$ radiation from photopion secondaries.

Fig.\ 2 shows a fit to the UHECR spectrum measured with the Pierre
Auger Observatory \cite{pao10} using 
a GRB model \cite{wda04} with different star formation rate factors
for GRBs \cite{wda04,hb06,led07}. If UHECRs originate from 
long duration GRBs, then the fit shown means that long GRBs 
have an absolute energy output of $\approx 6 (20)\times 10^{52}$ erg/GRB in 
cosmic rays, using a local long GRB rate density of $40$ 
Gpc$^{-3}$ yr$^{-1}$ \cite{gpw05} ($\approx 10$ 
Gpc$^{-3}$ yr$^{-1}$ \cite{led07}). This implies 
a factor $\sim 40$ -- $200\times$ more energy in 
cosmic rays than photons, using the photon emissivity of long-duration 
GRBs equal to $4\times 10^{43}$ erg Mpc$^{-3}$ 
yr$^{-1}\cong 4\times 10^{51}$ erg $\times 10$ Gpc$^{-3}$ yr$^{-1}$ \cite{led07}.
The efficiency for cosmic-ray generation must be even greater if the emissivity in 
UHECRs is compared only with nonthermal $\gamma$ rays at LAT energies \cite{eic10}.
Note, however, that these conclusions are specific to UHECR protons:
whereas the Auger team claims shower profiles are more consistent with heavy composition
at $\gtrsim 10^{19}$ eV \cite{aug09}, this claim is disputed by the HiRes Collaboration \cite{abb10}.
The intergalactic magnetic field must be strong, $\sim$ nG, to spread the arrival times of UHECR protons
\cite{rdf10}. 
%The recent claim \cite{ak10} of a low $\sim 10^{15}$G IGM field from {\it Fermi} LAT blazar data
%would threaten the hypothesis that GRBs accelerate UHECRs \cite{wda04}.
A galactic GRB component could make $\lesssim 10^{18}$ eV cosmic rays \cite{wda04,ckn10}. 
In terms of local luminosity density, an origin in BL Lac/FR I 
radio galaxies seems favored over GRBs, though GRBs have more than 
adequate power to accelerate UHECRs through {\it Fermi} processes \cite{dr10}. 

\section{Summary}

The briefest possible summary of a few of the interesting GRB results
from observations with {\it Fermi} has been given here. The goal of identifying 
hadronic signatures in the high-energy spectra of GRBs is ambiguous, and 
leptonic emission models are energetically favored. The {\it Fermi} LAT GRBs
show evidence for minimum bulk Lorentz factors $\gtrsim 10^3$, which 
gives interesting though not always energetically favorable implications
for hadronic models.

\vskip0.2in
\noindent I thank Katsuaki Asano, David Eichler, Kohta Murase, and Eli Waxman for discussions.
I would also like to thank S.\ Nagataki, K.\ Ioka, and the organizers for their kind invitation. 
The work of CDD is supported by the Office of Naval Research. 

\bibliographystyle{aipproc}   % if natbib is available
%\bibliographystyle{aipprocl} % if natbib is missing

%%%%%%%%%%%%%%%%%%%%%%%%%%%%%%%%%%%%%%%%%%%
%% You probably want to use your own bibtex database here
%%%%%%%%%%%%%%%%%%%%%%%%%%%%%%%%%%%%%%%%%%%
\bibliography{GRBbib}

%%%%%%%%%%%%%%%%%%%%%%%%%%%%%%%%%%%%%%%%%%%
%% Just a reminder that you may have to run bibtex
%% All of it up to \end{document} can be removed
%% if you don't like the warning.
%%%%%%%%%%%%%%%%%%%%%%%%%%%%%%%%%%%%%%%%%%%

%%%%%%%%%%%%%%%%%%%%%%%%%%%%%%%%%%%%%%%%%%%
%% The following lines show an example how to produce a bibliography
%% without the help of the BibTeX program. This could be used instead
%% of the above.
%%%%%%%%%%%%%%%%%%%%%%%%%%%%%%%%%%%%%%%%%%%

\end{document}